 \documentclass[12pt,preprint]{aastex}

\slugcomment{}

\shorttitle{Extinction Map of The Cygnus Region}
\shortauthors{Kohyama et al.}

\begin{document}


\title{A NEW GALACTIC EXTINCTION MAP OF THE CYGNUS REGION}


\author{T. Kohyama, H. Shibai and M. Fukagawa}
\affil{Graduate School of Science, Osaka University, 1-1 Machikaneyama-cho, Toyonaka, Osaka 560-0043, Japan}
\email{kohyama@iral.ess.sci.osaka-u.ac.jp}

\and

\author{Y. Hibi\altaffilmark{}}
\affil{Advanced Technology Center, National Astronomical Observatory of Japan, 2-21-1 Osawa, Mitaka, Tokyo 181-8588}



\begin{abstract}
We have made a Galactic extinction map of the Cygnus region with 5$'$ spatial resolution. The selected area is 80$^{\circ}$ to 90$^\circ$ in the Galactic longitude and -4$^\circ$ to 8$^\circ$ in the Galactic latitude. The intensity at 140 $\mu$m is derived from the intensities at 60 and 100 $\mu$m of the {\it IRAS} data using the tight correlation between 60, 100, and 140 $\mu$m found in the Galactic plane. The dust temperature and optical depth are calculated with 5$'$ resolution from the 140 and 100 $\mu$m intensity, and $A_{V}$ is calculated from the optical depth. In the selected area, the mean dust temperature is 17 K , the minimum is 16 K, and the maximum is 30 K. The mean $A_{V}$ is 6.5 mag, the minimum is 0.5 mag, and the maximum is 11 mag. The dust temperature distribution shows significant spatial variation on smaller scales down to 5$'$. Because the present study can trace the 5$'$-scale spatial variation of the extinction, it has an advantage over the previous studies, such as the one by Schlegel, Finkbeiner, \& Davis, who used the {\it COBE}/DIRBE data to derive the dust temperature distribution with a spatial resolution of 1$^\circ$. The difference of $A_{V}$ between our map and Schlegel et al.'s is $\pm$ 3 mag. A new extinction map of the entire sky can be produced by applying the present method. 

\end{abstract}


\keywords{dust, extinction --- infrared: ISM --- methods: data analysis}



\section{Introduction}

Radiation from extragalactic objects is generally reduced and attenuated the Galactic interstellar dust. As the dust extinction is high in the ultraviolet(UV) -- optical wavelength region, the observed data at these wavelengths must be corrected for the Galactic extinction along each line of sight. For example, statistical analyses of galaxies using the Sloan Digital Sky Survey (e.g., York et al.\ 2000) require an accurate extinction map. Schlegel, Finkbeiner, \& Davis (1998) (hereafter SFD98) have made an all-sky map of the Galactic extinction for this purpose by using the all-sky far-infrared maps of the Infrared Astronomy Satellite ({\it IRAS}) and the Diffuse Infrared Background Experiment (DIRBE) on the Cosmic Background Explorer ({\it COBE}). 

To derive extinctions, the dust size is important; it The dust size is classified into three groups (e.g., D\'esert et al.\ 1990): polycyclic aromatic hydrocarbon (PAH), very small grain (VSG), and large grain (LG). LG absorbs UV-optical photons and reradiates the absorbed energy in far infrared (FIR). The observed FIR emission is dominated by LGs (e.g., Sodroski et al.\ 1997). LG considered to be in equilibrium with the interstellar radiation field (ISRF) (e.g., Draine \& Anderson 1985) and radiates thermal emission with an equilibrium temperature. Therefore, assuming a spectral emissivity index, the dust temperature and the optical thickness can be derived from the FIR spectral energy distribution (SED). We can derive the dust extinction in the UV and optical wavelength regions from the FIR SED if the dust optical depth of LG is proportional to the extinction of each grain component (PAH+VSG+LG).

PAH and VSG, in contrast to LG, are heated stochastically by a single photon to high temperatures and cooled by radiation shorter than 60 $\mu$m. As a significant fraction of the 60 $\mu$m intensity usually comes from VSGs, we cannot derive the LG temperature from the {\it IRAS} 60 and 100 $\mu$m band data (e.g. D\'esert et al.\ 1990). 

Furthermore, there are two all-sky FIR diffuse maps obtained by ({\it IRAS}) and DIRBE. {\it IRAS} has two FIR photometric bands of 60 and 100 $\mu$m, while DIRBE has four FIR photometric bands of 60, 100, 140, and 240 $\mu$m. The spatial resolution of DIRBE is 0$^\circ$.7, whereas that of {\it IRAS} is 5$'$. 

SFD98 derived the dust temperature from DIRBE data at 100 and 240 $\mu$m, and published the all-sky extinction (reddening) map. They made an extinction map with 5$'$ resolution using the temperature derived from DIRBE and the 100 $\mu$m intensity of {\it IRAS}. Because the spatial resolution is limited by DIRBE, this extinction map may not trace small-scale variation as pointed out by Arce \& Goodman (1999).

Dobashi et al.\ (2005) made a visual extinction map with 6 $'$ resolution using the Digitized Sky Survey (DSS). Since the star counting technique cannot measure the dust behind the stars, its accuracy can worsen in directions where the extinction is high, depending on many parameters such as the angular density of stars, their location from the molecular cloud, and the distribution of the distances of the dust. Therefore, this technique may be ineffective for the correction of the Galactic extinction towards extragalactic objects near the Galactic plane.

To improve the maps, we assumed the color-color correlations between 60, 100, and 140 $\mu$m that were found by Hibi et al.\ (2006). They analyzed Zodi-Subtracted Mission Average (ZSMA) data of the Galactic plane within $\mid b \mid \ < 5^{\circ}$, LMC, and SMC. They identified two groups having very tight correlations in the color-color diagram. These correlations are not produced with the model by Li \& Draine (2002). Bot et al. (2009) also argued that the model by Draine \& Li (2007) does not explain the Multiband Imaging Photometer for Spitzer (MIPS) and {\it IRAS} data.

By using Hibi et al.'s correlation, we can obtain the dust temperature and the dust extinction with 5$'$ resolution only from the {\it IRAS} data. To test the validity of this method of making a high-resolution dust extinction map, we have analyzed the Cygnus region. In this paper, we report the results of our analysis and compare them with those of SFD98 and Dobashi et al.\ (2005). The data and the analysis method are shown in Section 2. The temperature and extinction maps are presented in Section 3. We discuss the results in Section 4. Section 5 presents to the summary.

\section{Data and Analysis}

%% In a manner similar to \objectname authors can provide links to dataset
%% hosted at participating data centers via the \dataset{} command.  The
%% second curly bracket argument is printed in the text while the first
%% parentheses argument serves as the valid data set identifier.  Large
%% lists of data set are best provided in a table (see Table 3 for an example).
%% Valid data set identifiers should be obtained from the data center that
%% is currently hosting the data.
%%
%% Note that AASTeX interprets everything between the curly braces in the 
%% macro as regular text, so any special characters, e.g. "#" or "_," must be 
%% preceded by a backslash. Otherwise, you will get a LaTeX error when you 
%% compile your manuscript.  Special characters do not 
%% need to be escaped in the optional, square-bracket argument.

In this section, the intensity at 140 $\mu$m is derived from the intensities at 60 and 100 $\mu$m of the {\it IRAS} data by using the tight correlation between 60, 100, and 140 $\mu$m found by Hibi et al.\ (2006). To test this method, we selected a sample area in the Cygnus region. The selected area is 80$^\circ$ to 90$^\circ$ in the Galactic longitude and $-4^\circ$ to $8^\circ$ in the Galactic latitude. The interplanetary dust (IPD) emission and the cosmic infrared background possibly contaminate the Galactic diffuse emission (e.g., Sodroski et al.\ 1997). However, as the Cygnus region is located at 60$^\circ$ in the ecliptic latitude, the subtraction of the IPD component has only a minor contribution to the uncertainty about our derivation of extinction discussed later. The cosmic infrared background emission is negligible, since the FIR intensity is high enough. 

In general, the interstellar space in the Galaxy is optically thin in the FIR wavelength region. The FIR intensity is written by
\begin{equation}
I_{\nu}({\lambda})=\tau_{100\ \mu m}\times\left(\frac{100\ \mu m}{\lambda}\right)^\beta \times B_{\nu}(\lambda, T_{d}).
\end{equation} 
In Equation (1), $I_{\nu}(\lambda)$ is the intensity at wavelength $\lambda$, $\tau_{100\ \mu m}$ is the optical depth at 100 $\mu$m, $T_{d}$ is the dust temperature, and $\beta$ is the spectral emissivity index. $B_{\nu}(\lambda, T_{d})$ is the Planck function. Hereafter, $I_{\nu}({\lambda})$ is described as $I(\lambda)$. $I(\lambda)$ is not color-corrected. The color-corrected intensity, $I_{C}(\lambda)$, is described below. We adopted a one-zone model, i.e., the dust temperature is assumed to be constant along each line of sight. We adopted $\beta = 2$ according to the previous study (Drain \& Lee 1984, Hibi et al.\ 2006, SFD98) for comparison. Some observational results show that $\beta$ changes depending on the physical status of the ISM (e.g., Reach et al.\ 1995 and del Burgo et al.\ 2003).

Because the published DIRBE all-bands intensities were derived by assuming a source spectrum, $\nu I(\lambda)$ = constant, the color correction should be applied to the observed data with the photometric bands of DIRBE. The color correction factor $K_{\lambda} \ (\beta,T_{d})$ is determined from the spectral emissivity index and the dust temperature, if the spectrum is written by Equation (1). The conversion from the observed (uncorrected) intensity to the color-corrected intensity is given by Equation (2) with the color correction factor
\begin{equation}
I_{C}(\lambda)=\frac{I(\lambda)}{K_{\lambda}\left(\beta,T_{d}\right)}.
\end{equation}
The color correction factor of DIRBE is quoted from the {\it COBE}/DIRBE Explanatory Supplement (1995). The color correction was made for the intensity at 140 and 100 $\mu$m, but not for that at 60 $\mu$m, because the SED shape around 60 $\mu$m cannot be described by Equation (1) (Hibi et al.\ 2006). The correction affects the intensities by a few percent and hence is not crucial for the degree of extinction derived later.

The color-color diagram was made from the DIRBE data in the Cygnus region using the same procedure as Hibi et al.\ (2006). Hibi et al.\ (2006) indicated that there were two tight correlations in the Galactic plane. The major group is called the main-correlation, and the minor group the sub-correlation. The main-correlation and the sub-correlation are fitted with Equations (3) and (4), respectively.
\begin{equation}
\frac{I_{C}\left(140\ \mu m\right)}{I_{C}\left(100\ \mu m\right)} = 0.65\left(\frac{I\left(60\ \mu m\right)}{I_{C}\left(100\ \mu m\right)}\right)^{-0.78}.
\end{equation}
\begin{equation}
\frac{I_{C}\left(140\ \mu m\right)}{I_{C}\left(100\ \mu m\right)} = 0.93\left(\frac{I\left(60\ \mu m\right)}{I_{C}\left(100\ \mu m\right)}\right)^{-0.56}.
\end{equation}
Hibi et al.\ (2006) and Hirashita et al.\ (2007) indicated that the main-correlation represents the lines of sight along which the ISRF is constant, and the sub-correlation represents the lines of sight along which the ISRF varies greatly. 

The color-color diagram of the Cygnus region is shown in Figure 1. In contrast to the Galactic plane, the data points satisfying $I(60\ \mu m)/I_{C}\ (100\ \mu m) < 0.28$ do not follow the main-correlation and other data points seem to follow the sub-correlation with a small offset. Therefore, we tried to fit the data points in two regions divided at $I(60\ \mu m)/I_{C}\ (100\ \mu m) = 0.28$. Least-squares fitting with two parameters of power and offset in the color-color diagram provided the following two relationships:

\begin{equation}
\frac{I_{C}\left(140\ \mu m\right)}{I_{C}(100\ \mu m)} = 0.73\left( \frac{I\left(60\ \mu m\right)}{I_{C}\left(100\ \mu m\right)}\right)^{-0.75}.
\end{equation}

\begin{equation}
\frac{I_{C}\left(140\ \mu m\right)}{I_{C}(100\ \mu m)} = 1.09\left( \frac{I\left(60\ \mu m\right)}{I_{C}\left(100\ \mu m\right)}\right)^{-0.44}.
\end{equation}

The coefficients of correlations of the former and latter cases are 0.7 and 0.8, respectively. There are 653 and 495 data points for the regions of $I(60\ \mu m)/I_{C}\ (100\ \mu m) < 0.28$ and $> 0.28$, respectively.

We calculate $I_{C}\ (140\ \mu m)$ only from $I(60\ \mu m)$ and $I_{C}\ (100\ \mu m)$ for four cases listed in Table 1 in addition to the SFD98 method. At first, we use Equation (6) for all data points, and call this the single case. As shown in Figure 1, this relation is similar to the sub-correlation, and approximately represents all the data points.

However, the scatter of the data points around the lines of Equations (5) and (6) does not reflect the noise included in the DIRBE's individual data points but actual variation. The noise level of DIRBE140 is 2 MJy/sr ({\it COBE}/DIRBE Explanatory Supplement 1995). Actually, the rms noise of DIRBE140 is 2 MJy/sr in the high latitude sky ($b > 70^{\circ}$). As the lowest intensity is 30 MJy/sr in the present analysis of the Cygnus region, the noise does not contribute to the scatter in Figure 1. The noise levels of DIRBE60 and DIRBE100 are negligible compared to the intensities at those bands. Therefore the scatter should be actual variation, the fitting with a single line as above introduces an additional error into the estimation of temperature. Therefore, we adopted the following method for the latter three cases. Each data point in Figure 1 was plotted from the DIRBE data with poor spatial resolution. SFD98 determined the dust temperature only from DIRBE100 and DIRBE140, and thus, small-scale variation of the dust temperature was smeared out. On the other hand, we can use IRAS60 and IRAS100 with better spatial resolution, and we are aware of the reasonably good correlations. Therefore, we initially plotted a single line with the same inclination as that of Equation (5) or (6) just at the data point plotted based on the DIRBE. Next, we apply the ratio of IRAS60 and IRAS100 to the vertical axis of Figure 1 to determine the ratio of IRAS100 intensity and the $140\ \mu m$ intensity. Equation (7) indicates this method. 

\begin{equation}
\frac{I_{C}\left(140\ \mu m\right)_{\ \ \ \ \ }}{I_{C}(100\ \mu m)_{IRAS}} = \left(\frac{I_{C}\left(140\ \mu m\right)}{I_{C}(100\ \mu m)}\right)_{DIRBE} \left( \left(\frac{I\left(60\ \mu m\right)}{I_{C}\left(100\ \mu m\right)} \right)_{IRAS} \div \left(\frac{I\left(60\ \mu m\right)}{I_{C}\left(100\ \mu m\right)}\right)_{DIRBE} \right)^{a}.
\end{equation}

The power $a$ is -0.75 and -0.44. 

According to the correlation recognized in Figure 1, we adopted three cases as in Table 1: inclination of Equation (5), inclination of Equation (6), and a combination of both inclinations. We call the case of $a = -0.75$ the slow case and the case of $a = -0.44$ the steep case. The last case is called the best-fit case that changes the power according to Equation (8) for better fitting;

\begin{equation}
a = 
\left\{
\begin{array}{cc}
-0.75 & \left( \left(\frac{I(60\ \mu m)}{I_{C}(100\ \mu m)}\right)_{IRAS} < 0.28\right) \\
-0.44 & \left( \left(\frac{I(60\ \mu m)}{I_{C}(100\ \mu m)}\right)_{IRAS} \geq 0.28\right).
\end{array}
\right .
\end{equation}

This new method, corresponding to the latter three cases, is based on the fact that the scatter around a single fitted line may represent some actual variation and there may still remain a correlation with a similar inclination around each data point. It can be interpreted that SFD98 adopted a vertical line at each data point for the derivation of the temperature in Figure 1, whereas the latter three cases of the present study adopted a tilted line at each data point representing the actual correlation found by Hibi et al. (2006). The scatter from Equation (5) or (6) produces no error. The DIRBE data were re-gridded from $0.32^{\circ}$ pixels to 5$'$ pixels, the same as those of {\it IRAS} with the sphere-approximation interpolation. As a result of this transformation, $I_{C} (140\ \mu m)$ of each pixel can be calculated using Equation (7).

%%\begin{equation}
%%\frac{I_{C}\left(140\ \mu m\right)}{I_{C}(100\ \mu m)} =
%%\left\{
%%\begin{array}{cc}
%%\left(\frac{1}{0.43}\frac{I\left(60\ \mu m\right)}{I_{C}\left(100\ \mu m\right)}\right)^{-1.30}. & \left(\frac{I(60\ \mu m)}{I_{C}(100\ \mu m)} < 0.25\right) \\
%%0.93\left(\frac{I\left(60\ \mu m\right)}{I_{C}\left(100\ \mu m\right)}\right)^{-0.56}+0.05. & \left(\frac{I(60\ \mu m)}{I_{C}(100\ \mu m)} > 0.25\right)
%%\end{array}
%%\right .
%%\end{equation}

Figure 2 shows the difference in the methods of calculating the dust temperature among five cases: SFD98, single, slow, steep, and best-fit. The best-fit is represented by the combination of the steep case and the slow case at $I(60\ \mu m)/I_{C} (100\ \mu m) = 0.28$. The difference between the present method and SFD98 appears in the difference between the inclinations of lines. The deviation of $I(60\ \mu m)/I_{C} (100\ \mu m) = 0.28$ of the {\it IRAS} data from that of the DIRBE data is reflected for $I_{C}\ (140\ \mu m)/I_{C} (100\ \mu m) = 0.28$ that corresponds to the dust temperature. Therefore, small-scale variation of the dust temperature should arise between the present method and SFD98.

In the above mentioned method, the $140\ \mu m$ intensity was calculated from the 60 and $100\ \mu m$ intensities of the {\it IRAS} data with DIRBE data for the four cases. The Improved Reprocessing of the {\it IRAS} Survey (IRIS) by Miville \& Lagache (2005) was used as the {\it IRAS} image. The IRIS maps are IPD-subtracted ones, the calibration is rescaled with the DIRBE absolute calibration, and the strip patterns are eliminated with a 2-D Fourier filter.

The {\it IRAS} data must be color-corrected as the DIRBE data. The color-correction factors were quoted from the {\it IRAS} Explanatory Supplement (Beichman et al. 1988). The dust temperature was calculated from the observed $100\ \mu m$ intensity and the derived $140\ \mu m$ intensity by Equations (6) and (7). The optical depth at $100\ \mu m$ is written as
\begin{equation}
\tau_{100\ \mu m}=\frac{I_{C}(100\ \mu m)}{B_{\nu}(100\ \mu m,T_{d})}.
\end{equation}
$A_{V}$ is calculated by
\begin{equation}
A_{V}=865.85\times A_{100\ \mu m}=891\times\tau_{100\ \mu m}.
\end{equation}
The value of the conversion factor from $\tau_{100\ \mu m}$ to $A_{V}$ is quoted from Mathis (1990) by assuming $R_{V} = 3.1$. This value decreases by 15 \% in the case of $R_{V} = 5.0$.

The notation for $A_{V}$ is summarized in Table 1.

\begin{table}[htbp]
\begin{center}
\caption[Table 1]{\centering Calculation methods for SFD98 and four new cases. \label{Table 1}}
\begin{tabular}{ll}
\hline
\hline
Notation & Method \\
\hline
$A_{V}$ (SFD98) & SFD98 \\
$A_{V}$ (single) & A single best-fit line, Equation (6)\\
$A_{V}$ (slow) & A line with the inclination of Equation (5) at each DIRBE data point\\
$A_{V}$ (steep) & Same as $A_{V}$ (slow) but with the inclination of Equation (6)\\
$A_{V}$ (best) & Combination of $A_{V}$ (slow) and $A_{V}$ (steep) of Equation (8)\\
\hline
\end{tabular}
\end{center}
\end{table}

%% In this section, we use  the \subsection command to set off
%% a subsection.  \footnote is used to insert a footnote to the text.


%% This section also includes several of the displayed math environments
%% mentioned in the Author Guide.

\section{Result}

$A_{V}$ values in the Cygnus region are calculated for the four cases. Figure 3 shows the histogram of $A_{V}$ (best) $-$ $A_{V}$ (SFD98). The standard deviation is 3.4. Figure 4 shows the histograms of $A_{V}$ (best) $-$ $A_{V}$ (slow) and $A_{V}$ (best) $-$ $A_{V}$ (steep). The standard deviations are 1.2 and 0.83, respectively. Figure 5 shows the histogram of $A_{V}$ (steep) $-$ $A_{V}$ (single). The standard deviation is 1.9.

Figure 6 compares the dust temperature maps made by the best-fit case of our study with those by SFD98. The dust temperature distribution of the present study shows small-scale spatial variation that is not apparent in the map by SFD98. The maximum and the standard deviation of the temperature difference between the two maps are 5 K and 0.5 K, respectively. 

Figure 7 shows the $A_{V}$ distribution of this area. The fluctuation of both the maps seems to have similar angular scales. However, the $A_{V}$ values of both the maps differ from each other.

To analyze the reason for these differences, the differences in temperature and $A_{V}$ are shown in Figure 8. The sign of the left panel of Figure 8 is reversed compared to that of the right panel so as to visually check the dependence because $A_{V}$ has a negative dependence on temperature. It can be seen that the two maps resemble each other remarkably, which means that the difference in $A_{V}$ originates from the difference in the derived temperature. 

\section{Discussion}

\subsection{Comparison with SFD98}

The methods using the best-fit and steep cases are considered to be more precise than the SFD98 method because of the following reason. As described in Section 3, the $A_{V}$ map derived by the best-fit differs from that by SFD98, and the difference can be ascribed to the fact that the present temperature map has higher spatial resolution compared with that of SFD98. Figure 9 shows the result of comparison between the $A_{V}$ values of the best-fit case and SFD98. The difference, ($A_{V}$ (best) $-$ $A_{V}$ (SFD98))/$A_{V}$ (best), scatters by 21 \% in 1 sigma. SFD98 employed the {\it IRAS} $100\ \mu m$ intensity to calculate the extinction. They removed point sources from the {\it IRAS} $100\ \mu m$ map to smoothen the map with a FWHM = 3.2$'$ Gaussian profile. The difference in the $100\ \mu m$ intensity between our map and that of SFD98 is 9 \% (1 sigma) in the Cygnus region. This difference can also account for the $A_{V}$ difference between the present study and SFD98 (see Equations (9) and (10) ) in addition to the difference between spatial resolution of dust temperature; the scatter seen in Figure 9 is affected by this difference. The dust temperature difference in the $A_{V}$ difference is estimated as 19 \% by Equation (11);

\begin{equation}
\Delta A_{V}(T_{d}) = \sqrt{ \Delta A_{V}\ ^{2}(total) - \Delta A_{V}\ ^{2}(I (100\ \mu m)}.
\end{equation}

$\Delta A_{V}(T_{d})$, $\Delta A_{V}(total)$, and $\Delta A_{V}(I (100\ \mu m))$ is the dust temperature difference at the $A_{V}$ difference, total $A_{V}$ difference (21 \%), and the $100\ \mu m$ intensity difference at $A_{V}$ difference (9 \%), respectively. On the other hand, the noise of the IRIS data is 0.03 MJy/sr for $\lambda = 60\ \mu m$ and 0.06 MJy/sr for $\lambda = 100\ \mu m$ (Miville \& Lagache 2005). These values correspond to less than 1 \% for $A_{V}$ and are negligibly small compared to the difference between $A_{V}$ (best) and $A_{V}$ (SFD98).

Figure 10 shows the comparison between $A_{V}$ (best) and $A_{V}$ (slow). It can be seen that the difference is large compared to the noise of the DIRBE data points, reflecting the fact that the power of the slow case does not represent the entire correlation in the Cygnus region. Figure 11 shows the comparison between $A_{V}$ (best) and $A_{V}$ (steep). The difference is 5 \% (1 sigma of ($A_{V}$ (best) $-$ $A_{V}$ (steep))/$A_{V}$ (best)). The difference between $A_{V}$ (steep) and $A_{V}$ (SFD98) is 20 \% (1 sigma of ($A_{V}$ (steep) $-$ $A_{V}$ (SFD98))/$A_{V}$ (steep)). Thus, the dust temperature difference in the $A_{V}$ difference is estimated as 18 \% by Equation (11). Figure 12 shows the comparison of $A_{V}$ (steep) and $A_{V}$ (single). The difference, ($A_{V}$ (steep) $-$ $A_{V}$ (single))/$A_{V}$ (steep)), scatters by 21 \% in 1 sigma. It is consistent with the hypothesis described in Section 2 that the steep, slow, and best-fit cases are more accurate compared with the single case. 
 
\subsection{Comparison with the Star Counting Method}

Dobashi et al. (2005) published an $A_{V}$ map within $\mid b \mid \ < 40^\circ$ by the star counting method using the DSS images. The spatial resolution is 6$'$, similar to the present study. Figure 13 shows the result of the comparison of both $A_{V}$ data in the Cygnus region. The $A_{V}$ (DSS) data is distributed lower than $A_{V}$ (best) and saturated at $A_{V}$ (DSS) $>$ 5 mag. This indicates that the star counting method with DSS has a limit of around $A_{V}$ (DSS) $\sim$ 5 mag. It is apparent that their method is not useful for $A_{V}$ $>$ 10 mag because the optical radiation does not reach us because of heavy extinction by the dust. Dobashi (2009) applied their star counting method to the Two-Micron All Sky Survey (2MASS) to reach $A_{V}$ $\sim$ 30 mag.

\subsection{Selecting Valid Correlation}

As shown in Section 4.1, $A_{V}$ by the best-fit and steep cases are more precise than that by SFD98 in the Cygnus region. But these methods would not be effective in high Galactic latitudes because $I_{C}\ (140\ \mu m)$ has a lower S/N ratio in those latitudes. Therefore, if we choose only one case out of the best-fit, slow, steep, the main-correlation, and the sub-correlation for the entire sky, the sub-correlation is best suited because of its similarity to the steep case.

In the entire sky, there are regions where both the main-correlation and the sub-correlation can be seen (e.g., the Galactic plane). There are two methods for determining whether the main- or sub-correlation is applicable if the best-fit is not used. The first method is the same as that for the Cygnus region. We can make a color-color diagram in any area of the sky and choose the better one. The second method involves referring to the radio thermal continuum map that represents the bulk of the ionized gas excited by the OB stars. The direction with strong radio continuum should be associated with the region where the ISRF varies on a small scale because of the strong local heating sources. Therefore, the FIR colors in the direction with strong radio continuum can be thought to follow the sub-correlation as demonstrated by Hibi et al. (2006). Figure 1 shows that the sub-correlation is more major than the main-correlation in the Cygnus region.

We estimate the systematic difference arising from the selection of either the main-correlation or the sub-correlation, since in general we are unaware a priori which of the two correlations is applicable to a certain region. We compare $A_{V}$ calculated by the main-correlation and the sub-correlation in Figure 14. The range of the $A_{V}$ ratio is between 1 and 2.5 where the range of $I (60\ \mu m)/I_{C} (100\ \mu m)$ is between 0.2 and 0.6.

\subsection{Validity of Applying Correlations to {the \it IRAS} Map}

Hibi et al. (2006) identified the two groups in the color-color plot of the DIRBE data with 42$'$ resolution. Each group has good correlation expressed by Equations (3) and (4). These two expressions can be applied to the intensity distribution with a higher spatial resolution such as that of the {\it IRAS} map. As indicated by Hibi et al. (2006) and Hirashita et al. (2007), the main-correlation represents the lines of sight along which the ISRF is constant, and the sub-correlation represents the lines of sight along which the ISRF varies greatly. Obviously, the lines of sight showing the sub-correlation may also represent the direction along which the ISRF varies within the beam. 
The direction following the main-correlation with the DIRBE beam does not include regions having strong radiation (Case 1). Otherwise, it must follow the sub-correlation. The ISRF does not vary extensively along the line of sight and over the spatial extent of the DIRBE beam. Therefore, all lines of sight of the {\it IRAS} beam in a DIRBE beam should follow the main-correlation. 

On the other hand, there should be strong ISRF regions in the line of sight indicating the sub-correlation with the DIRBE beam (Case 2). The strong ISRF region is usually compact, so the spatial extent could be smaller than the DIRBE beam. A DIRBE beam can be divided into many {\it IRAS} beams, some indicating the main-correlation (Case 2-1) and the others the sub-correlation (Case 2-2). The former does not include the strong ISRF regions, and the latter includes the strong radiation field region in the line of sight. 

In Case 2-1, the difference in the derived $A_{V}$ between the main-correlation and the sub-correlation must be small because there is only a small difference between the main-correlation and the sub-correlation under weak ISRF. If the sub-correlation is applied, the maximum error of $A_{V}$ is 30 \%, where $I (60\ \mu m)/I_{C} (100\ \mu m)$ is lower than 0.3 (Fig. 14).

Finally, we consider Case 2-2. The {\it IRAS} data points that include a strong radiation field in the DIRBE beam also follow the sub-correlation, because such a strong radiation field does not entirely occupy the line of sight. Therefore, there is no error in using the sub-correlation in Case 2-2.

Consequently, even if a higher spatial resolution map (e.g., {\it IRAS}) is employed, the $A_{V}$ error does not increase to a large extent.

\subsection{Future Prospect}

An all-sky extinction map can be produced using the present method. In higher Galactic latitudes, extinction is generally small and so is the temperature fluctuation. Then the $A_{V}$ difference between the present study and SFD98 would be smaller in the Cygnus region. The subtraction of the IPD thermal emission is an important issue for high Galactic latitudes or low ecliptic latitudes. The {\it AKARI} all-sky diffuse map will be released at 60, 100, 140, and 160 $\mu$m with better spatial resolution in the future. Then, the present study should be compared with the {\it AKARI} map.

\section{Summary}
We have derived the $A_{V}$ map with 5$'$ resolution in the Cygnus region using the color-color correlation between 60, 100, and 140 $\mu$m discovered by Hibi et al. (2006). The $A_{V}$ difference between the present study and SFD98 is significantly larger than the uncertainty associated with our derivation of the intensity at $140 \mu m$ with 5$'$ resolution. The difference may occur because the spatial resolution of the present temperature map is ten times higher than that of SFD98. Therefore, the present method is more accurate than that of SFD98. An all-sky extinction map can be produced by the present method. 


%% Included in this acknowledgments section are examples of the
%% AASTeX hypertext markup commands. Use \url without the optional [HREF]
%% argument when you want to print the url directly in the text. Otherwise,
%% use either \url or \anchor, with the HREF as the first argument and the
%% text to be printed in the second.

\acknowledgments

We are grateful to Dr. H. Hirashita for comments on the draft and useful discussions. This work was supported by Grant-in-Aid for challenging Exploratory Research (22654023).

\clearpage

%% the file, uncomment the \plotone, \plottwo, and
%% If you need a layout that cannot be achieved with \plotone or
%% \plottwo, you can invoke the graphicx package directly with the
%% \includegraphics command or use \plotfiddle. For more information,
%% please see the tutorial on "Using Electronic Art with AASTeX" in the
%% documentation section at the AASTeX Web site,
%% http://www.journals.uchicago.edu/AAS/AASTeX.
%%
%% The examples below also include sample markup for submission of
%% supplemental electronic materials. As always, be sure to check
%% the instructions to authors for the journal you are submitting to
%% for specific submissions guidelines as they vary from

%% This example uses \plotone to include an EPS file scaled to
%% 80% of its natural size with \epsscale. Its caption
%% has been written to indicate that additional figure parts will be
%% available in the electronic journal.

\begin{figure}
\begin{center}
\includegraphics[width=13cm,bb=70 150 580 680, clip]{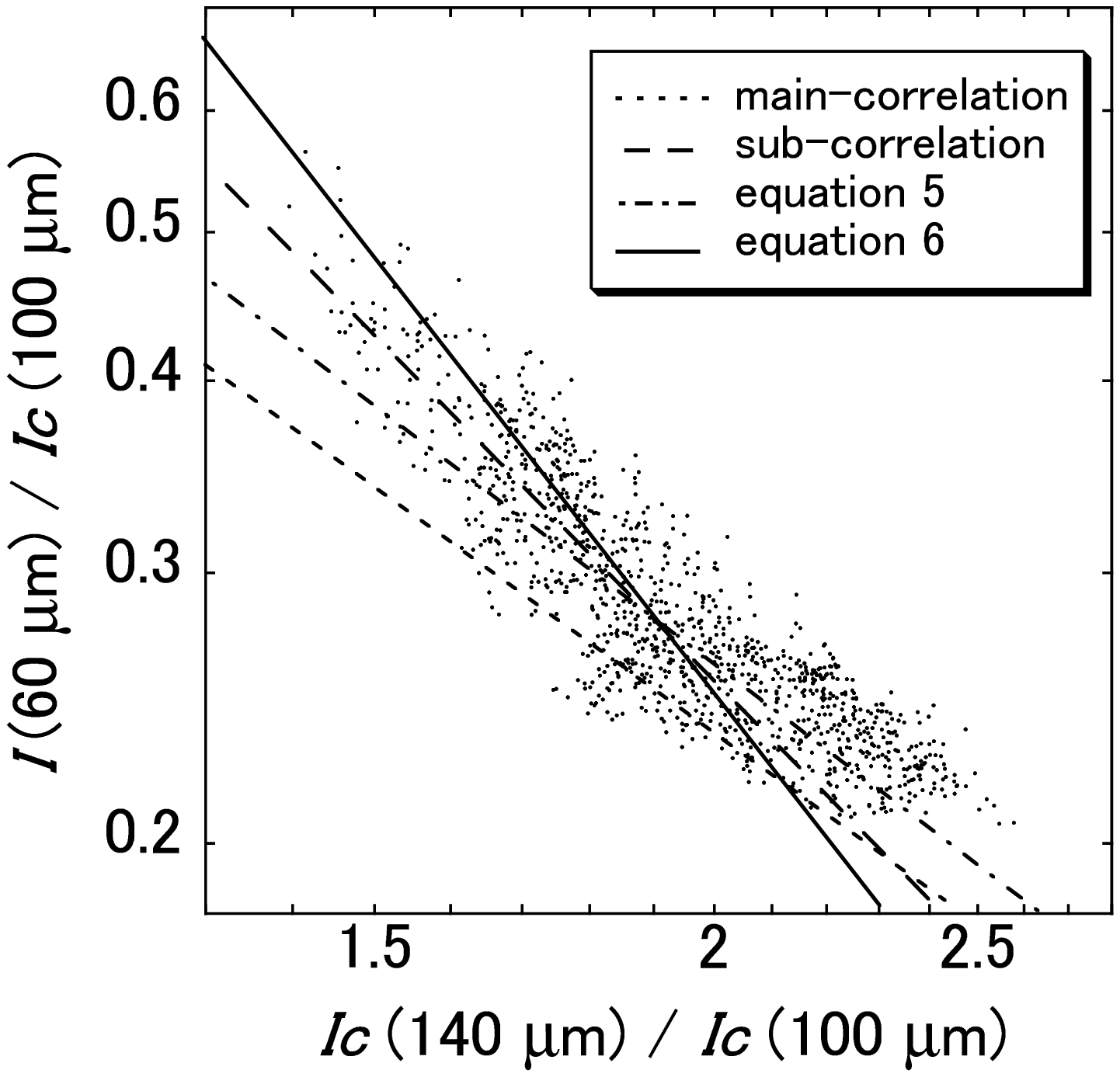}
\end{center}
\caption{Color-color diagram between 60, 100, and 140 $\mu$m in the Cygnus region. The horizontal axis is the 140 -100 $\mu$m color, while the vertical axis is the 60 -100 $\mu$m color. The data points are from DIRBE. The main-correlation, the sub-correlation, Equation (5), and Equation (6) are represented by the doted line, dashed line, dashed-dotted line, and solid line, respectively.\label{fig1}}
\end{figure}

\clearpage

\begin{figure}
\begin{center}
\includegraphics[width=13cm,bb=70 150 580 680, clip]{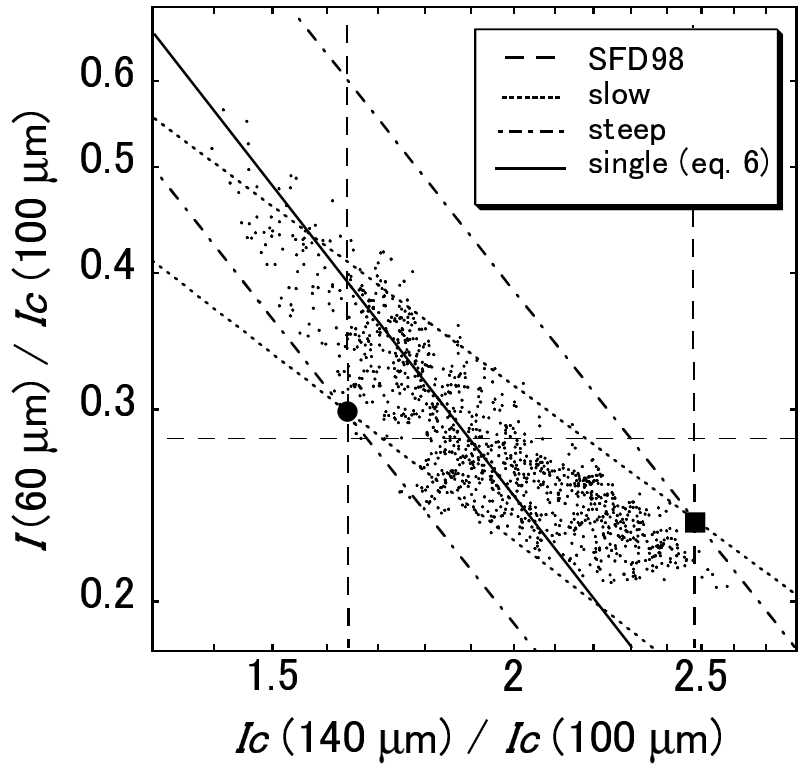}
\end{center}
\caption{Explanation of $A_{V}$ calculation on the color-color diagram between 60, 100, and 140 $\mu$m in the Cygnus region. The horizontal dashed line shows $I (60\ \mu m)/I_{C}\ (100\ \mu m) = 0.28$. The circle and square are points arbitrarily selected in regions of $I (60\ \mu m)/I_{C} (100\ \mu m) > 0.28$ and $<$ 0.28, respectively.\label{fig2}}
\end{figure}

\clearpage

\begin{figure}
\begin{center}
\includegraphics[width=13cm,bb=60 160 570 670,clip]{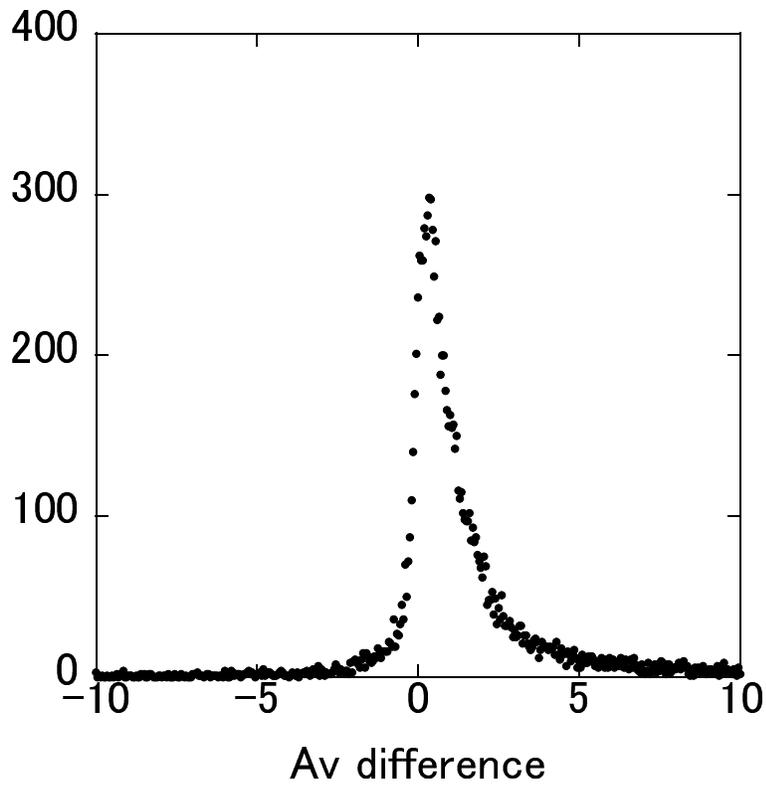}
\caption{Histogram of the $A_{V}$ difference of $A_{V} (best) - A_{V}(SFD98)$. Bin size is 0.05.\label{fig3}}
\end{center}
\end{figure}

\clearpage

\begin{figure}
\begin{center}
\includegraphics[width=13cm,bb=60 160 570 670,clip]{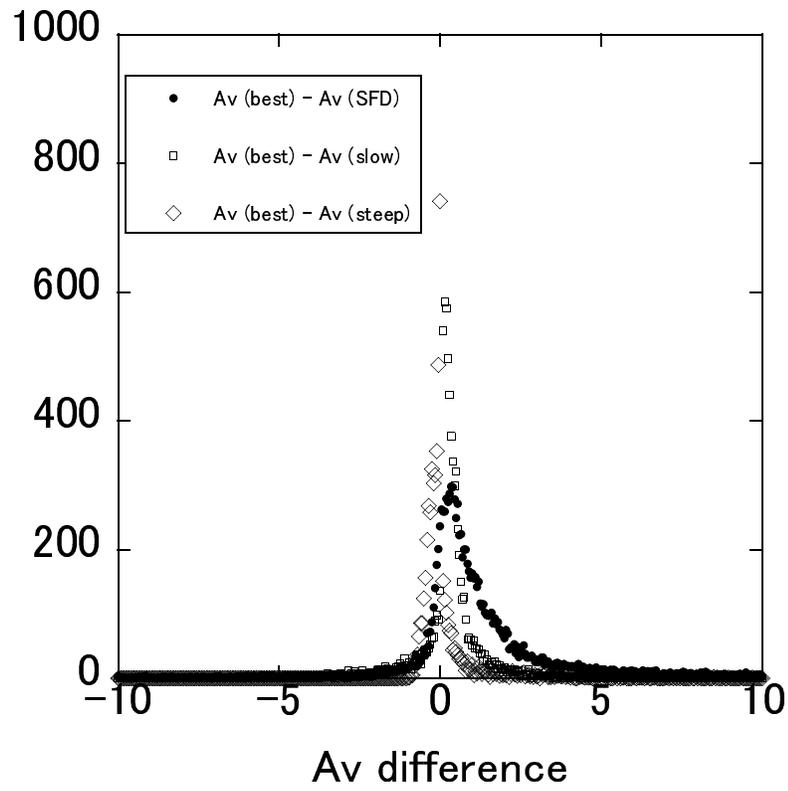}
\caption{Histogram of $A_{V}$ differences. Bin size is 0.05. Filled circles represent the same difference as in Fig. 2.\label{fig4}}
\end{center}
\end{figure}

\clearpage

\begin{figure}
\begin{center}
\includegraphics[width=13cm,bb=60 160 570 670,clip]{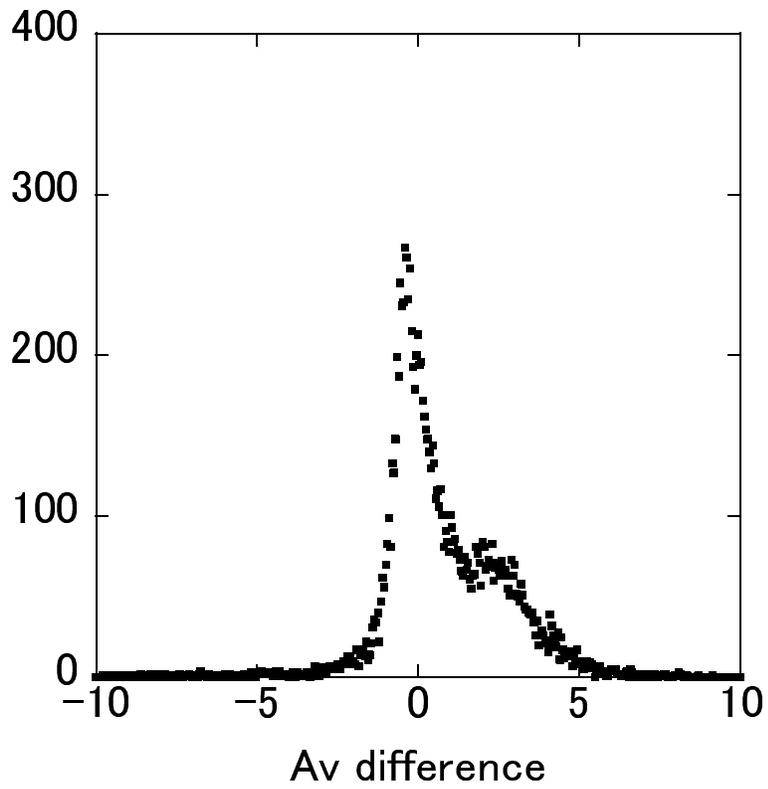}
\caption{Histogram of $A_{V}$ difference of $A_{V} (steep) - A_{V}(single)$. Bin size is 0.05.\label{fig5}}
\end{center}
\end{figure}

\clearpage

\begin{figure}
\begin{center}
\includegraphics[width=17cm,bb=60 270 800 800,clip]{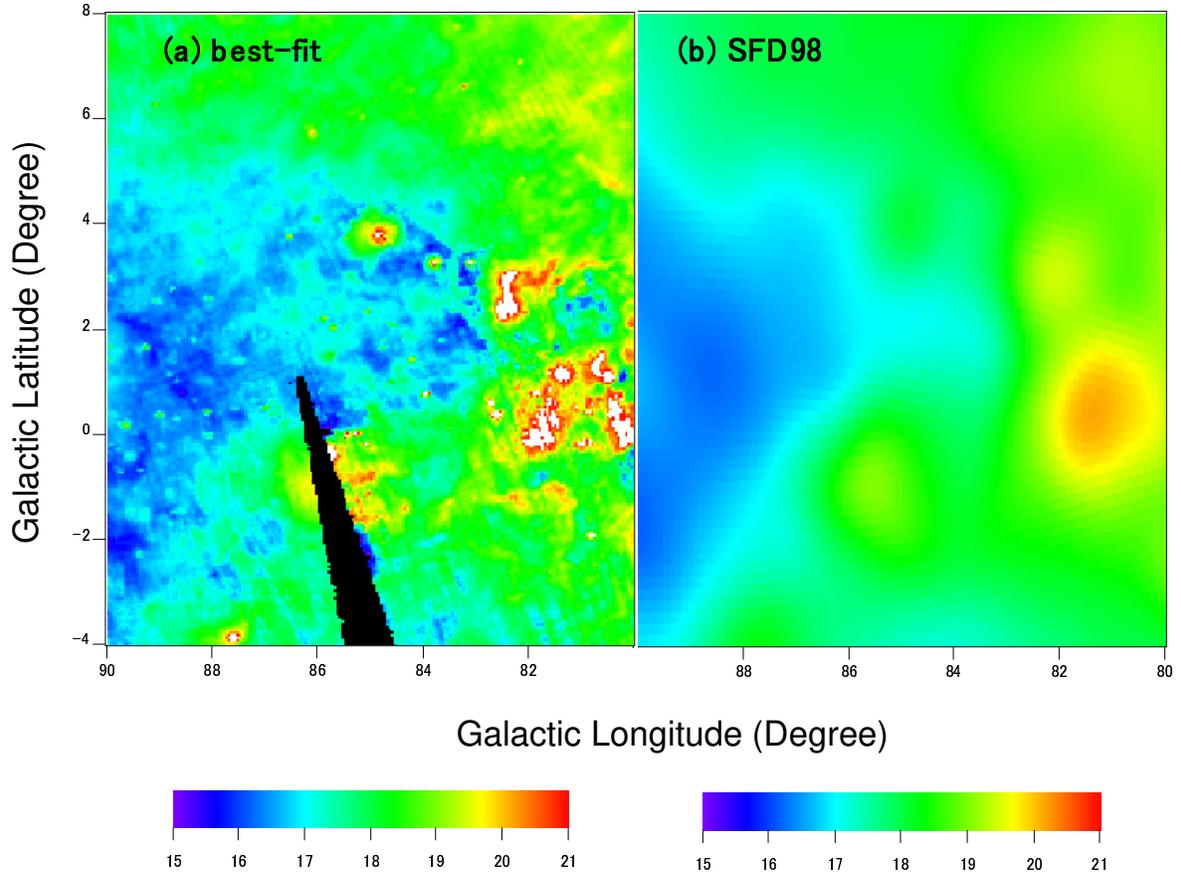}
\caption{Dust temperature distribution maps in the Cygnus region: (a) present study and (b) that by SFD98. The black area in Panel (a) indicates the area that {\it IRAS} did not survey.\label{fig6}}
\end{center}
\end{figure}

\clearpage

\begin{figure}
\begin{center}
\includegraphics[width=17cm,bb=60 270 800 800,clip]{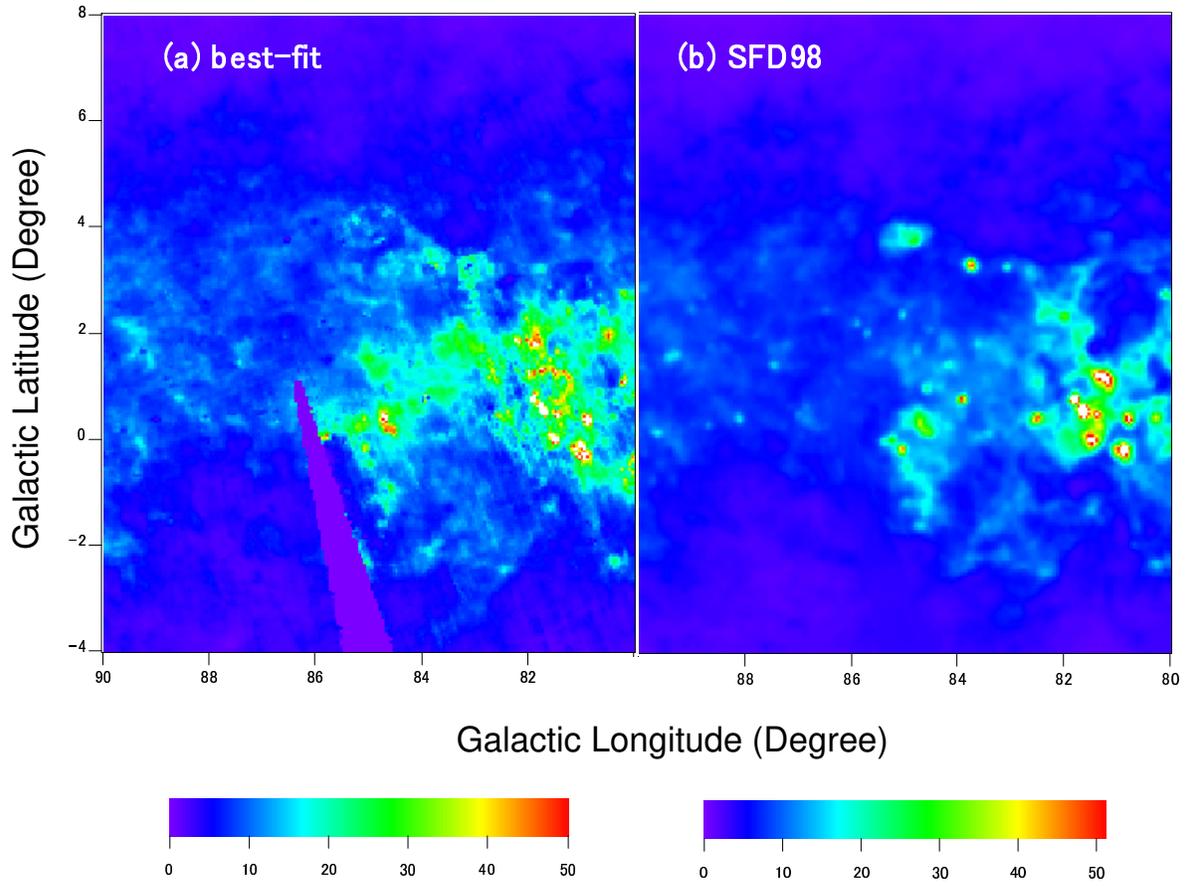}
\caption{$A_{V}$ distribution maps: (a) present study and (b) that by SFD98.\label{fig7}}
\end{center}
\end{figure}

\clearpage

\begin{figure}
\begin{center}
\includegraphics[width=17cm,bb=60 270 800 800,clip]{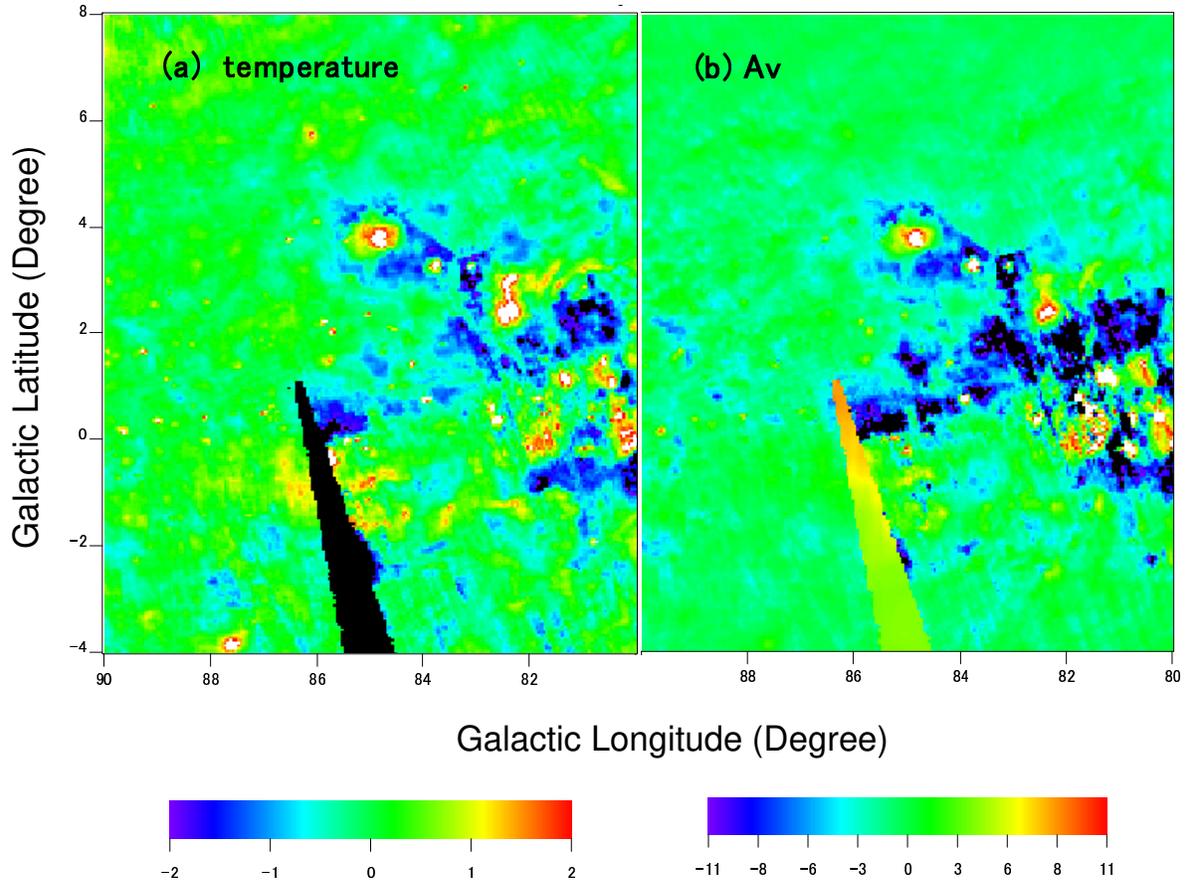}
\caption{Difference maps in temperature (a) and in $A_{V}$ (b). The sign is reversed for the left panel.\label{fig8}}
\end{center}
\end{figure}

\clearpage

\begin{figure}
\begin{center}
\includegraphics[width=13cm,bb =60 170 570 675,clip]{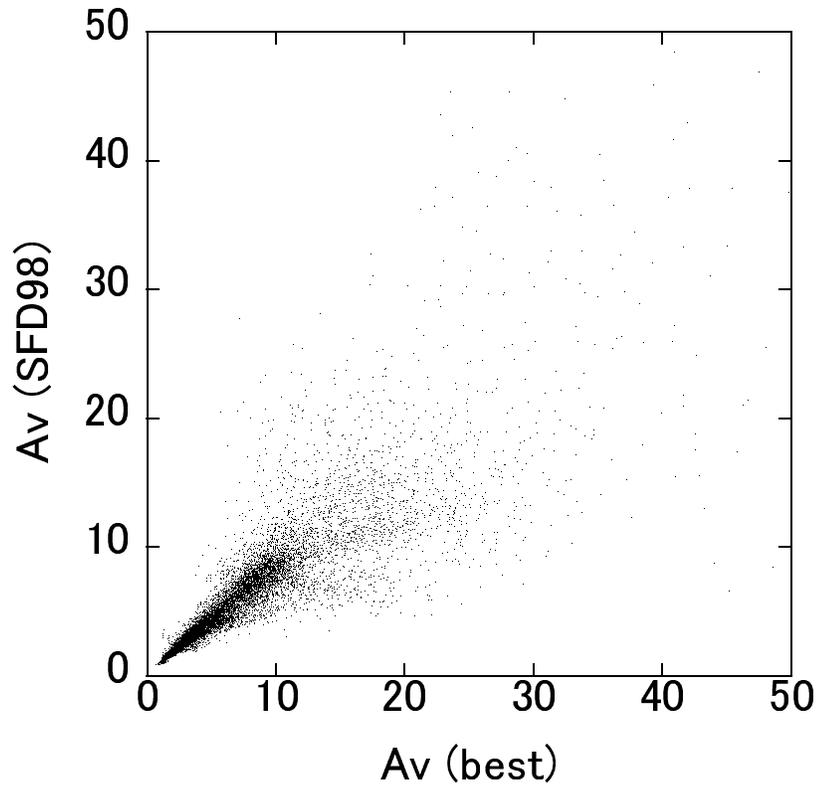}
\caption{Comparison between the $A_{V}$ values of the best-fit case and SFD98.\label{fig9}}
\end{center}
\end{figure}

\clearpage

\begin{figure}
\begin{center}
\includegraphics[width=13cm,bb=60 150 570 655,clip]{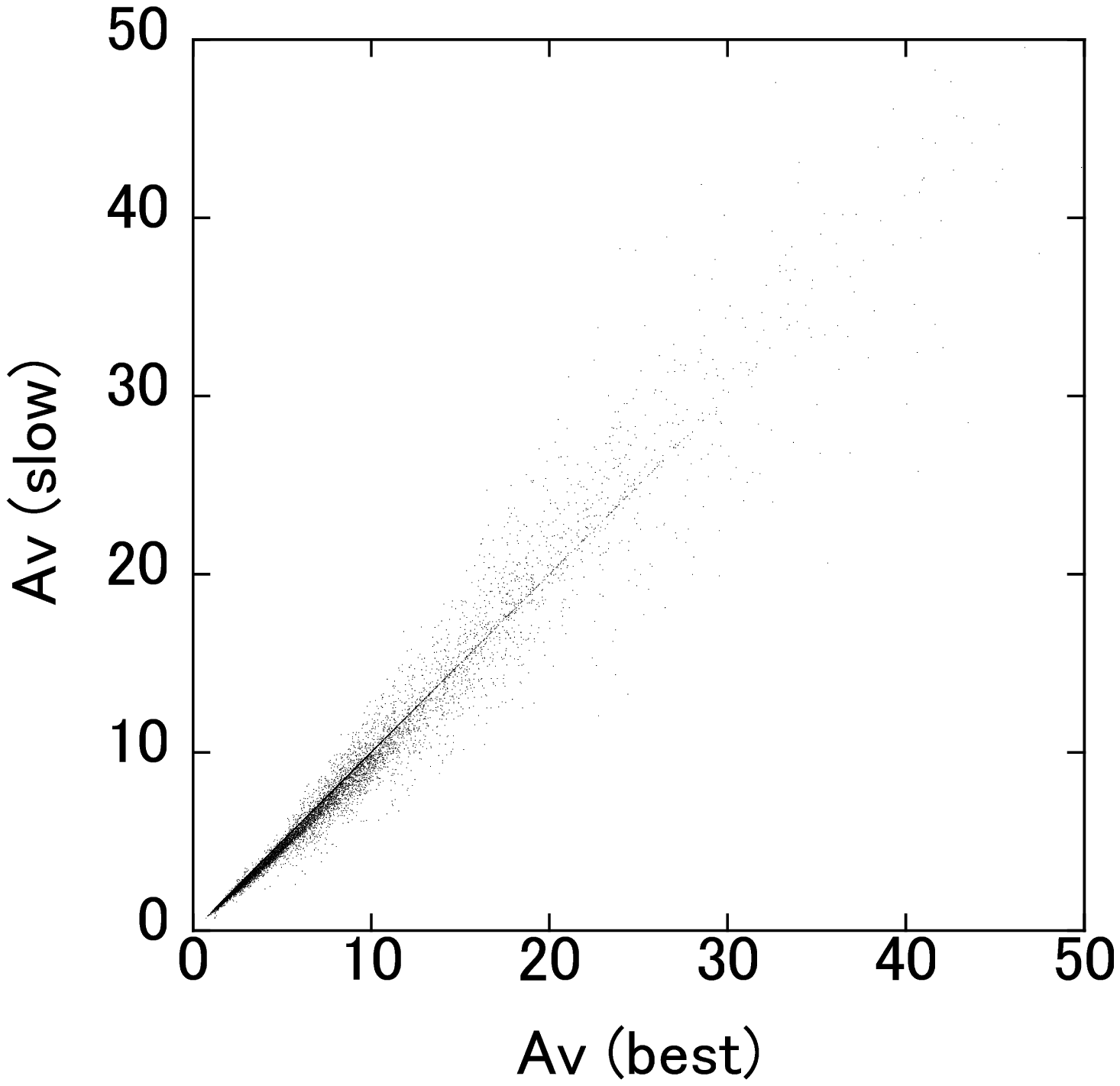}
\caption{Comparison between the $A_{V}$ values of the best-fit case and the slow case.\label{fig10}}
\end{center}
\end{figure}

\clearpage

\begin{figure}
\begin{center}
\includegraphics[width=13cm,bb=60 150 570 655,clip]{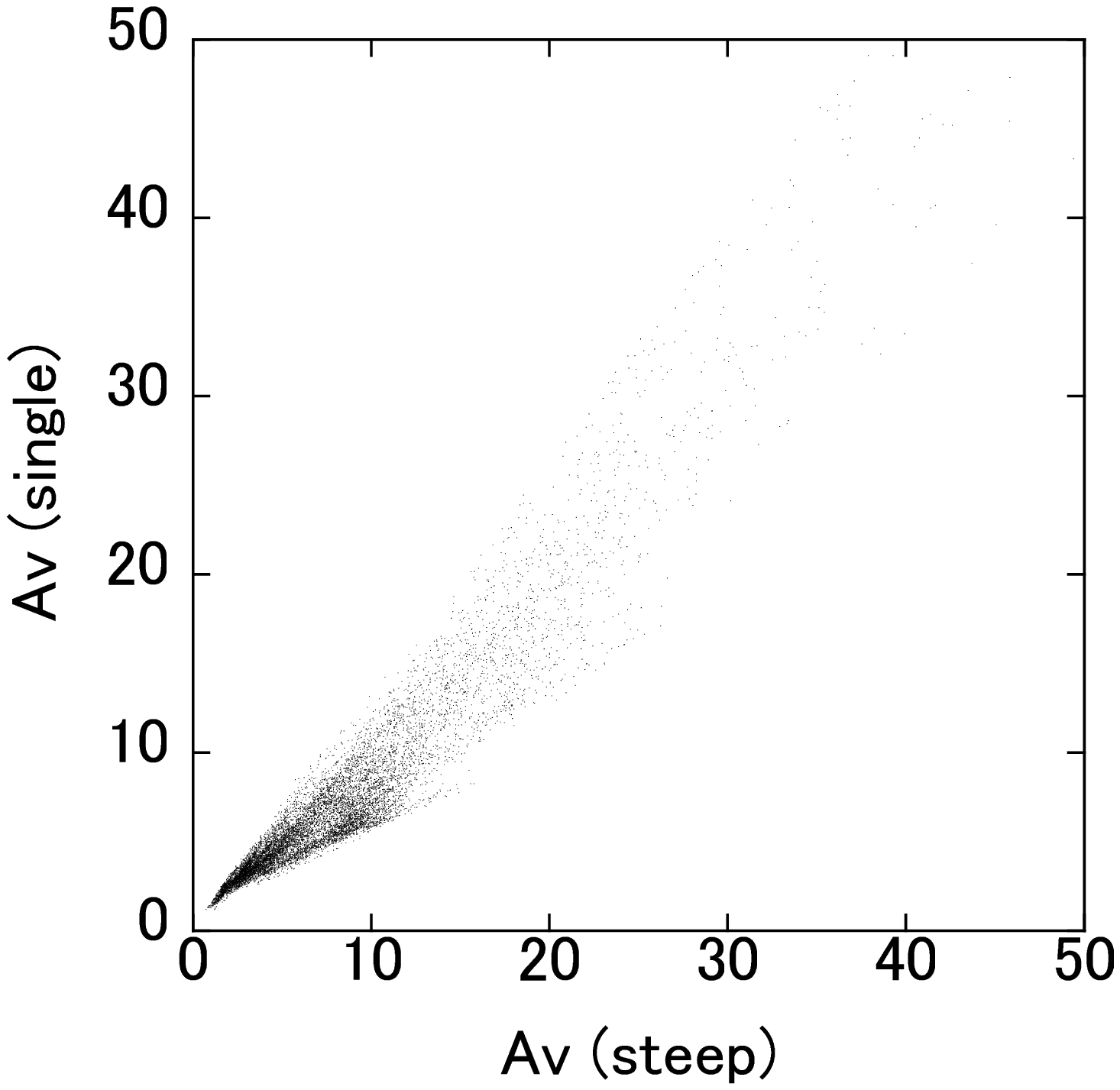}
\caption{Comparison between the $A_{V}$ values of the best-fit case and the steep case.\label{fig11}}
\end{center}
\end{figure}

\clearpage

\begin{figure}
\begin{center}
\includegraphics[width=13cm,bb=60 150 570 655,clip]{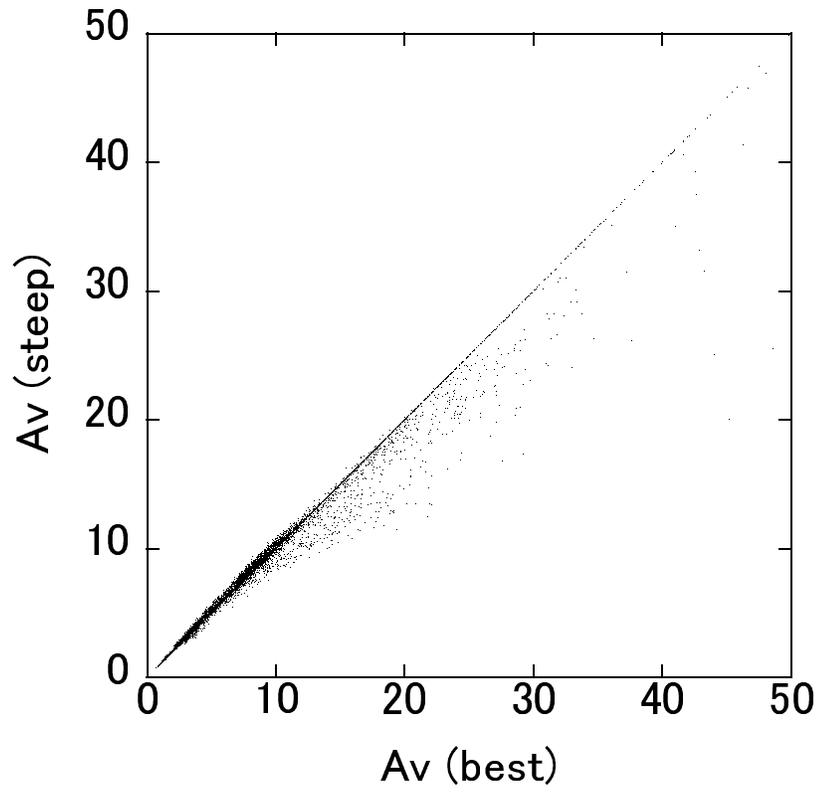}
\caption{Comparison between the $A_{V}$ values of the steep case and the single case.\label{fig12}}
\end{center}
\end{figure}

\clearpage

\begin{figure}
\begin{center}
\includegraphics[width=13cm,bb=60 180 570 620,clip]{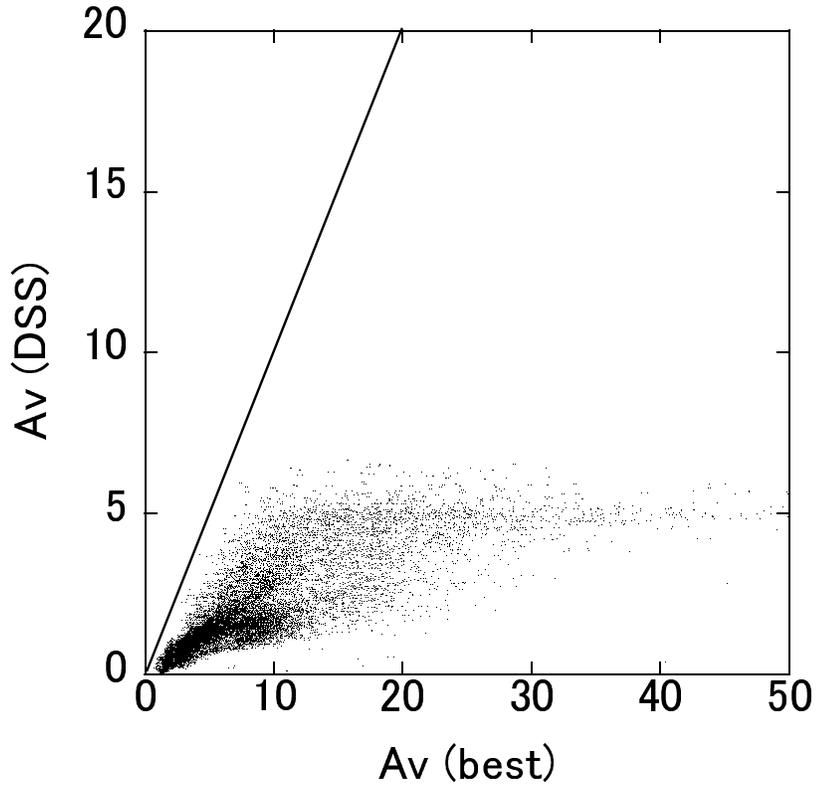}
\caption{Comparison between the $A_{V}$ values of the best-fit case and Dobashi et al. (2005). The solid line indicates a slope of unity.\label{fig13}}
\end{center}
\end{figure}

\clearpage

\begin{figure}
\begin{center}
\includegraphics[width=13cm,bb=60 150 570 700,clip]{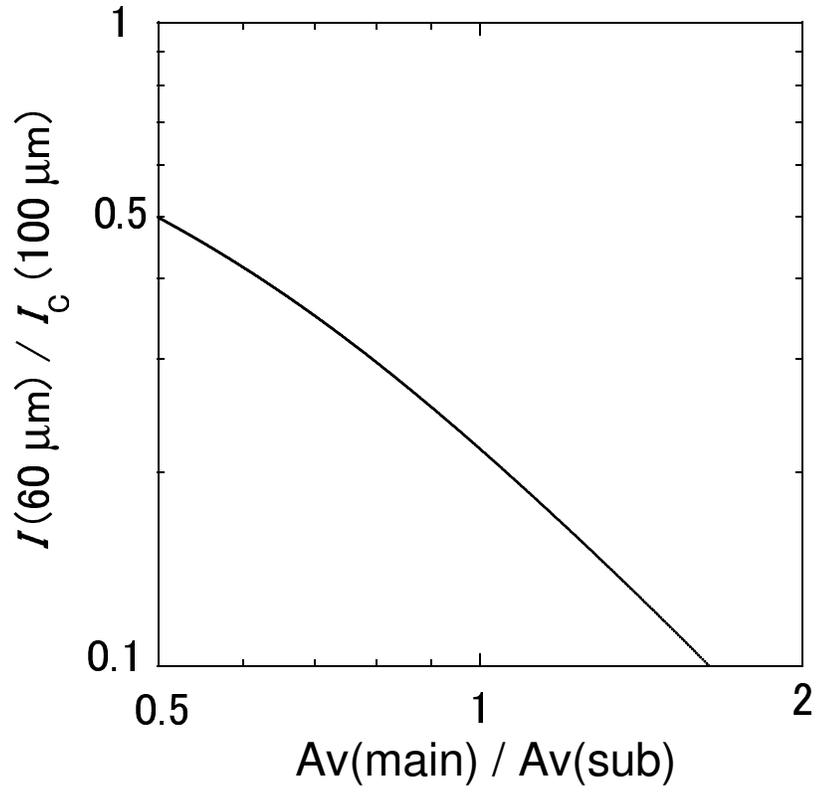}
\caption{Uncertainty of $A_{V}$ when the sub-correlation is adopted. The solid curve shows the ratio of $A_{V}$ by the main-correlation over the sub-correlation. \label{fig14}}
\end{center}
\end{figure}

\clearpage

\clearpage
\end{document}